\documentclass[aps,prb,amsmath,amssymb,reprint,superscriptaddress,preprintnumbers,showpacs,intlimits,longbibliography]{revtex4-1}
\usepackage{bm,latexsym,mathrsfs,enumerate}
\usepackage[table,dvipsnames]{xcolor}

\usepackage[breaklinks=true,unicode=true,urlcolor = blue,colorlinks = true,citecolor = blue,linkcolor = blue]{hyperref}
\usepackage{graphicx}
\usepackage{mathtools}
\usepackage{longtable}
\usepackage{multirow}
\usepackage{makecell,tabularx}
\usepackage{hhline}
\graphicspath{{./figs/}}
\renewcommand{\vec}[1]{\bm{#1}}

\bibpunct{[}{]}{,}{n}{}{}

\usepackage{array}
\newcolumntype{C}[1]{>{\centering\let\newline\\\arraybackslash\hspace{0pt}}m{#1}}
\newcolumntype{q}[1]{>{\centering}m{#1}}
\begin{document}
\title{Micromagnetic Theory of Curvilinear Ferromagnetic Shells}

\author{Denis D. Sheka}
\email{sheka@knu.ua}
\affiliation{Taras Shevchenko National University of Kyiv, 01601 Kyiv, Ukraine}

\author{Oleksandr V. Pylypovskyi}
\email{engraver@knu.ua}
\affiliation{Taras Shevchenko National University of Kyiv, 01601 Kyiv, Ukraine}

\author{Pedro~Landeros}
\email{pedro.landeros@usm.cl}
\affiliation{Departamento de F\'{\i}sica, Universidad T\'{e}cnica Federico Santa Mar\'{\i}a, Avenida Espa\~{n}a 1680, Valpara\'{\i}so, Chile}
\affiliation{Center for the Development of Nanoscience and Nanotechnology (CEDENNA), 917-0124 Santiago, Chile}

\author{Yuri~Gaididei}
\email{ybg@bitp.kiev.ua}
\affiliation{Bogolyubov Institute for Theoretical Physics of National Academy of Sciences of Ukraine, 03680 Kyiv, Ukraine}

\author{Attila K\'akay}
\email{a.kakay@hzdr.de}
\affiliation{Helmholtz-Zentrum Dresden-Rossendorf e.V., Institute of Ion Beam Physics and Materials Research, 01328 Dresden, Germany}

\author{Denys Makarov}
\email{d.makarov@hzdr.de}
\affiliation{Helmholtz-Zentrum Dresden-Rossendorf e.V., Institute of Ion Beam Physics and Materials Research, 01328 Dresden, Germany}

%%%%%%%%%%%%%%%%%%%%%%%%%%%%%%%%%%%%%%%%%%%%%%%%%%%%%%%%%%%%%%%%%%%%%70
%
%         ABSTRACT
%
%%%%%%%%%%%%%%%%%%%%%%%%%%%%%%%%%%%%%%%%%%%%%%%%%%%%%%%%%%%%%%%%%%%%%70

\begin{abstract}
	Here, we present a micromagnetic theory of curvilinear ferromagnets, which allows discovering novel fundamental physical effects which were amiss. In spite of the firm confidence for more than 70 years, we demonstrate that there is an intimate coupling between volume and surface magnetostatic charges. Evenmore, the physics of curvilinear systems requires existence of a new fundamental magnetostatic charge determined by local characteristics of the surface. As a stark consequence, novel physical nonlocal anisotropy and chiral effects emerge in spatially corrugated magnetic thin films. Besides these fundamental discoveries, this work reassures confidence in theoretical predictions for experimental explorations and novel devices, based on curved thin films.
\end{abstract}

%\pacs{75.75.-c, 75.78.-n, 75.78.Jp, 75.78.Cd}

% 75.75.-c	  Magnetic properties of nanostructures
% 75.78.-n	  Magnetization dynamics
% 75.78.Jp    Ultrafast magnetization dynamics and switching
% 75.78.Cd    Micromagnetic simulations

%%%%%%%%%%%%%%%%%%%%%%%%%%%%%%%%%%%%%%%%%%%%%%%%%%%%%%%%%%%%%%%%%%%%%70

\maketitle

%%%%%%%%%%%%%%%%%%%%%%%%%%%%%%%%%%%%%%%%%%%%%%%%%%
%%%			New section
%%%%%%%%%%%%%%%%%%%%%%%%%%%%%%%%%%%%%%%%%%%%%%%%%%

\section{Introduction}
\label{sec:intro}

Physical properties of living \cite{McMahon05} but also synthetic systems in condensed \cite{Castelvecchi17} and soft \cite{Senyuk12} matter are determined by the interplay between the physical order parameter, geometry and topology. Specifically to magnetism, magnetization textures and dynamic responses become sensitive to bends and twists in physical space. 
Curvature effects emerged as a novel tool in various areas of physics to tailor electromagnetic properties and responses relying on geometrical deformations  \cite{Bowick09,Turner10}. Typically, the consideration of curvature-induced effects is based on theories, which involve \emph{local} interactions for the description of molecular alignment in liquid crystals \cite{Napoli12,Napoli13,Napoli18}, physics of superconductors \cite{Vitelli04,Gladilin08,Fomin12}, macromolecular structures \cite{Forgan11} electronic properties of different corrugated films \cite{Cortijo07,Juan11,Yan13a,Gentile15,Ochoa17}.  For many systems, if not for all, this local description is incomplete. For instance, in magnetically ordered systems local picture misses to describe most of micromagnetic textures like chiral domain walls, skyrmion-bubbles and vortices. Therefore, the accepted fundamental foundation of modern magnetism necessarily requires both local and nonlocal interactions threatement on equal footing~\cite{Landau35,Brown63,Aharoni96,Hubert09}.

In contrast, the modern theory of curvilinear magnetism is still at the level when local~\cite{Carvalho-Santos08,Kravchuk12a,Gaididei14,Sheka15, Pylypovskyi15b,Sheka15c,Yershov15b,Pylypovskyi16, Yershov16,Moreno17a,Gaididei18a,Volkov18,Kravchuk18a} and non-local~\cite{Landeros07,Landeros10,Sheka13b,Sloika14, Yershov15,Otalora16, Hertel16,Tretiakov17,Otalora17, Sloika17,Otalora18} interactions are treated separately. This makes the description of the systems inherently incomplete as not only important fundamental effects can be amiss but also predictive power of the available theory is limited.

Here, we present a generalized micromagnetic theory of curvilinear magnetism. The theory describes the impact of curvature induced effects, driven by both local and nonlocal interactions, on static and dynamic magnetic texture in curved magnetic thin shells. Fundamentally, we identified new effects, which do no exist in planar magnets. In particular, we demonstrated that the physics of curvilinear systems cannot be described in the frame of the established physical picture relying on surface and volume magnetostatics charges, introduced in a seminal work by W.~F.~\citet{Brown63}. The curvature leads to the appearance of the new magnetostatic charge, determined by local characteristics of the surface. This newcomer is responsible for the appearance of novel fundamental effects like nonlocal anisotropy and nonlocal chiral effects. Furthermore, for more than 70 years there was a firm confidence that the surface and volume magnetostatic charges are decoupled. They were always considered as the two sides of the same coin. We demonstrate that there appears an intimate coupling between these two quantities. As a stark consequence, novel chiral effects emerge in spatially corrugated magnetic thin films.

These new effects are completely unexplored. We are convinced that their analysis will stimulate to rethink the origin of chiral and anisotropy effects in different systems, e.g. in fundamentally appealing and technologically relevant skyrmionic systems in polycrystalline thin films where surface roughness is unavoidable.

On the technical side, we apply a novel mathematical framework based on covariant derivatives formalism which allows to separate explicit curvature effects from spurious effects of the curvilinear reference frame. Relying on symmetry consideration of different interactions we predict and classify possible curvature effects on a equilibrium state magnetic texture in curved magnetic thin shells, which goes beyond the well-accepted linear in film thickness approach. 

The impact of this theory goes well beyond the magnetism community. The presented conclusions can be easily extended for studying the evolution of generic vector fields on curved shells in different models of condensed (graphene \cite{Yan13a}, superconductors \cite{Vitelli04}) and soft (nematics \cite{Napoli12}, cholesterics \cite{Napoli13}) matter.

%%%%%%%%%%%%%%%%%%%%%%%%%%%%%%%%%%%%%%%%%%%%%%%%%%
%%%			New section
%%%%%%%%%%%%%%%%%%%%%%%%%%%%%%%%%%%%%%%%%%%%%%%%%%

\section{Results}
\label{sec:results}

%==================================================================\
\begin{figure*}%[h]
	\includegraphics[width=\textwidth]{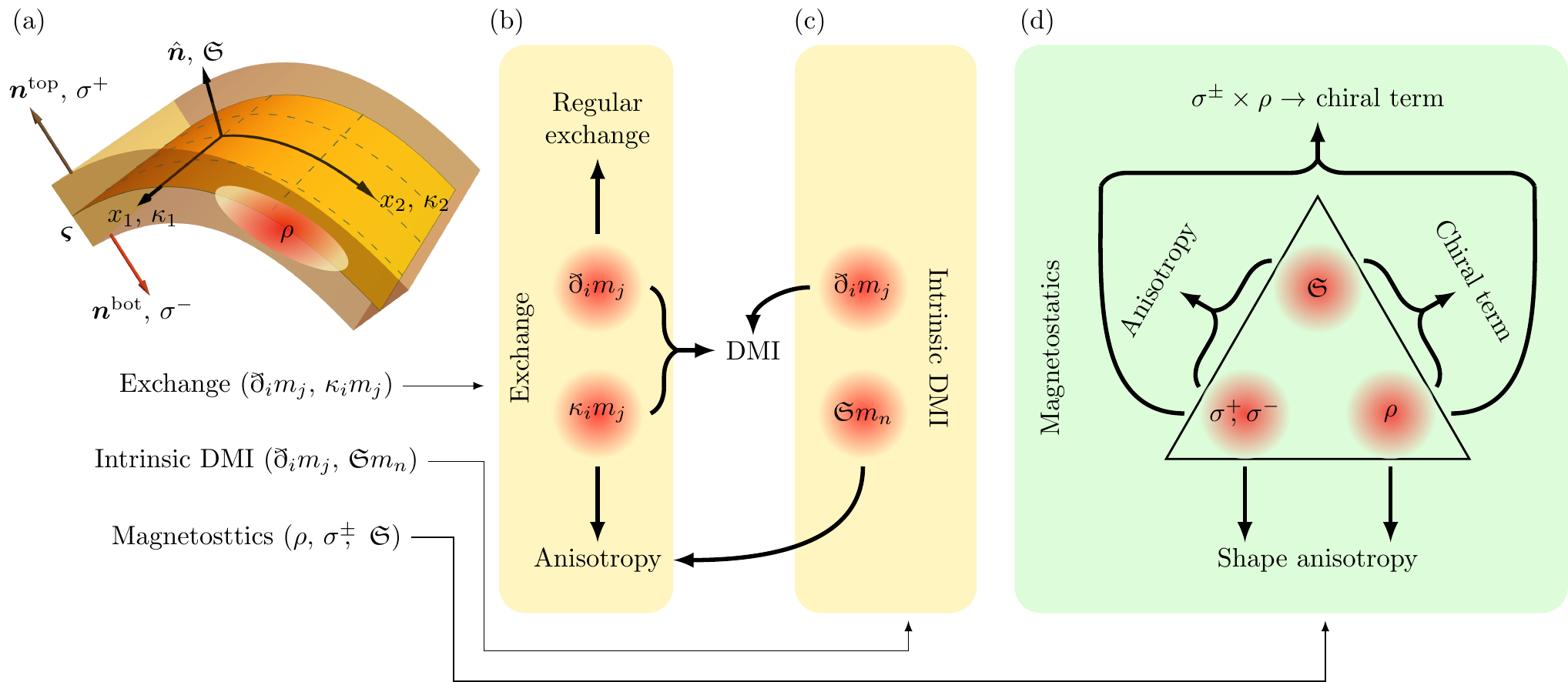}
	\caption{(Color online) \textbf{Impact of curvature on ferromagnetic shell.} (a) Schematics of surface with corresponding Darboux three-frame. Principal directions are shown by dashed lines. Central surface $\vec{\varsigma}$ is extruded along the normal direction $\hat{\vec{n}}$. Top and bottom surfaces have their own normals $\vec{n}_\textsc{s}^\text{top}$ and $\vec{n}_\textsc{s}^\text{bot}$, respectively. (b) \textit{Exchange} interaction is locally defined by tangential derivatives of magnetization $\eth_im_j$ and curvature-driven terms $\kappa_i m_j$. (c) \textit{Intrinsic DMI} is locally defined by $\eth_im_j$ as well as exchange, and by mean curvature-driven term $\mathfrak{S}m_n$. Exchange and intrinsic DMI both contribute to the regular exchange term and chiral term of interfacial DMI type in the total energy as well, as anisotropy of uni- and biaxial type. (d) \textit{Magnetostatics} in general form is expressed through three type of charges: volume $\rho$, surface $\sigma^\pm$ and cuvature-induced one $\mathfrak{S}$. Their combination results in shape anisotropy (in the same way as for flat samples), anisotropic and two chiral terms.
	}
	\label{fig:schematics}
\end{figure*}
%==================================================================/

We consider a curved ferromagnetic shell of thickness $h$ with a shape locally described by a Gaussian $\mathcal{K}(\vec{r})$ and mean $\mathcal{H}(\vec{r})$ curvatures. A  magnetic texture is controlled by the exchange, anisotropy and magnetostatic interactions. The total energy, normalized by $4\pi M_s^2$, has the following form:
\begin{equation} \label{eq:E-total}
\begin{aligned}
E &= E_{\text{x}} + E_{\text{an}} + E_{\text{d}}, \\
E_{\text{x}} &=- \ell^2 \int \mathrm{d} \vec{r}\, \vec{m}\cdot \vec{\nabla}^2\vec{m},\\
E_{\text{an}} &=\frac{Q}{2} \int \mathrm{d} \vec{r} \left(\vec{m} \cdot \vec{e}_a\right)^2, \\
E_{\text{d}} &=\frac{1}{8\pi} \!\!\int\!\! \mathrm{d} \vec{r} \!\! \int \!\! \mathrm{d} {\vec{r}'} \left(\vec{m}(\vec{r}) \!\cdot\! \vec{\nabla}\right)\left(\vec{m}({\vec{r}'}) \!\cdot\! {\vec{\nabla}'}\right) \! \frac{1}{\left|\vec{r} - {\vec{r}'}\right|},\!\!\!\! \\
E_{\textsc{dm}} & = D \int \mathrm{d} \vec{r} \left( m_n \nabla \vec{m} - (\vec{m}\cdot \nabla) m_n \right).
\end{aligned}
\end{equation}
Here, $\ell = \sqrt{A/4\pi M_s^2}$ is the exchange length, $A$ is the exchange constant, $M_s$ is the saturation magnetization, $Q = K/(2\pi M_s^2)$ with $K$ being the intrinsic crystalline anisotropy constant, $\vec{e}_a = \vec{e}_a(\vec{r})$ is the direction of the anisotropy axis and $\vec{m}(\vec{r}) = \vec{M}/M_s$ is the unit vector of magnetization. We suppose that the anisotropy direction $\vec{e}_a$ is determined by the surface geometry, it corresponds to one of the principal directions or their linear combination, see Appendix \ref{sec:geometry} for details, $D = D_\text{int}/(4\pi M_s^2)$ with $D_\text{int}$ being constant of the intrinsic Dzyaloshinskii--Moriya interaction (DMI), and $m_n$ is the normal component of magnetization.

We limit our discussion to the case of thin shells and describe them as an extrusion of a surface $\vec{\varsigma}(\vec{r})$ by a constant value $h$ along the vector $\vec{\hat{n}}= \vec{\hat{n}}(\vec{r})$ normal to the surface. Furthermore, we assume that the magnetization does not depend on the thickness coordinate along $\vec{\hat{n}}$.
By choosing the curvilinear reference frame, adapted to the geometry of an object, anisotropy obtains its usual spatially-invariant form, see Fig.~\ref{fig:schematics}.

For the theoretical analysis, we apply a new mathematical tool based on covariant derivatives. The main purpose of using this language is to separate two effects: an system-specific curvature effect and spurious effect of the curvilinear reference frame, see Appendix~\ref{sec:geometry} for details. Because of the geometry broken symmetry it is natural to restructure all magnetic energy terms containing spatial derivatives. A characteristic example is an exchange interaction: being isotropic in the Cartesian reference frame, it contains three components of different symmetries in curvilinear coordinates, $E_{\text{x}} = E_{\text{x}}^0  + E_{\text{x}}^\textsc{a} + E_{\text{x}}^\textsc{d}$~\cite{Gaididei14,Sheka15}. $E_{\text{x}}^0$ is a `common', regular isotropic part of exchange interaction, which has the form similar to the one in a planar film:
\begin{subequations} \label{eq:E-ex}
\begin{equation} \label{eq:E-ex-0}
E_{\text{x}}^0 = 
h \ell^2 \int \left(\eth_\alpha m_i\right) \left(\eth_\alpha m_i\right) \mathrm{d} S 
\end{equation}
with $\eth_\alpha$ being the modified covariant derivative with respect to the surface coordinate $x_\alpha$, see Eq.~\eqref{eq:D-covar}, and $m_i$ being the magnetization components in the curvilinear orthonormal Darboux three-frame $\left\{\vec{e}_1, \vec{e}_2, \hat{\vec{n}}\right\}$ on the surface $\vec{\varsigma}$, where $\vec{e}_1$ and $\vec{e}_2$ are unit vectors corresponding to the principal directions, $\hat{\vec{n}} = \vec{e}_1\times \vec{e}_2$ is the normal to the surface, see Fig.~\ref{fig:schematics} and Appendix \ref{sec:geometry} for details. Here and below, we use Greek letters $\alpha,\,\beta,\ldots = {1,2}$ to denote indices restricted to the shell surface. To indicate all three components of any vector, we use Latin indices $i,\,j,\ldots = {1,2,n}$.  Here and below we also use the Einstein summation convention. We emphasize, that all effects stem from the choice of the reference frame are properly and unambiguously assigned to $E_\text{ex}^0$. This represents the major advantage of the approach based on covariant derivatives.

The second term in the exchange energy reads
\begin{equation} \label{eq:Eex-A}
E_{\text{x}}^\textsc{a} =  h\ell^2 \int w_\text{x}^{\textsc{a}} \mathrm{d}S , \quad w_\text{x}^{\textsc{a}} = \mathscr{K}_{ij} m_im_j.  \\
\end{equation}
In general, this energy term describes the curvature-induced biaxial anisotropy, $\begin{Vmatrix}\mathscr{K}_{ij}\end{Vmatrix} = \text{diag} \left(\kappa_1^2, \kappa_2^2 ,\kappa_1^2 + \kappa_2^2\right)$ with $\kappa_1$ and $\kappa_2$ being local values of principle curvatures, related to Gaussian and mean curvature as $\mathcal{K} = \kappa_1 \kappa_2$ and $\mathcal{H} = \kappa_1 + \kappa_2$, respectively. Then, energy density of this term reads
\begin{equation}\label{eq:wex-a}
w_\text{x}^{\textsc{a}} = \kappa_1^2 m_1^2 + \kappa_2^2 m_2^2 + (\kappa_1^2 + \kappa_2^2)m_n^2.
\end{equation} 
A striking manifestation of the curvature-induced anisotropy is shape-induced patterning, for a review see \cite{Streubel16a}.

The last term in the exchange energy is a curvature-induced extrinsic DMI \cite{Gaididei14} 
\begin{equation} \label{eq:Eex-D}
\begin{aligned}
E_{\text{x}}^\textsc{d} = & 2h\ell^2 \int \mathrm{d}S \left(w_\text{x}^{\textsc{d}\,1} + w_\text{x}^{\textsc{d}\,2} \right)\\
& w_\text{x}^{\textsc{d}\,\alpha} = \kappa_\alpha \mathscr{L}_{\alpha n}^{(\alpha)},\quad \alpha=1,2,
\end{aligned}
\end{equation}
where no summation over $\alpha$ in Eq.~\eqref{eq:Eex-D} is applied. This term is
determined by the curvilinear-geometry analogue of Lifshitz invariants
\begin{equation} \label{eq:Lifshitz-inv}
\mathscr{L}_{i j}^{(\alpha)} = m_i \eth_{\alpha} m_j - m_j \eth_{\alpha} m_i.
\end{equation}
\end{subequations}
The two Lifshitz invariants in \eqref{eq:Eex-D} are determined by principal curvatures $\kappa_1$ and $\kappa_2$. The curvature-induced DMI is a reason for a chiral symmetry breaking, i.e. magnetochiral effects \cite{Hertel13a}, for a review see \cite{Streubel16a}.

The curvilinear geometry also has an affect on the magnetostatic energy of a shell $E_{\text{d}}$. For further analysis, it is insightful to modify magnetostatic volume charges:
\begin{equation} \label{eq:div-2D}
-\vec{\nabla}\cdot \vec{m} = -\eth_\alpha m_\alpha +\mathfrak{S}, \qquad \mathfrak{S}(\vec{r}) =  \mathcal{H}(\vec{r}) m_n(\vec{r}).
\end{equation}
Physics of curvilinear magnetism naturally introduces three fundamental charges: surface and volume magnetostatic charges introduced by \citet{Brown63}, and novel curvature-inducd charge $\mathfrak{S}$, determined by the mean curvature $\mathcal{H}$.
Although, the latter has a striking similarity to a `conventional' surface charge $\sigma^{\pm} = \vec{m}\cdot \vec{n}_{\textsc{s}}^\text{top~(bot)}$ on the top $+$ (bottom $-$) surface, which is also proportional to the normal component of the magnetization. Still, there is an important difference between $\mathfrak{S}$ and $\sigma^{\pm}$. The surface charges $\sigma^{\pm}$ have opposite signs at opposite shell surfaces. Hence, these charges act like an effective magnetostatic capacitor, see Fig.~\ref{fig:schematics}. In contrast, the curvature-induced charge $\mathfrak{S}$ is determined by the normal to the surface $\vec{\varsigma}$ (but not via the top/bottom surface of a shell). Furthermore, the sign of $\mathfrak{S}$ is defined by the  mean curvature $\mathcal{H}$ only. This new charge leads to the appearance of new physical effects which are intrinsically nonlocal and reveal themselves as nonlocal anistorpy and nonlocal chiral effects.

Similar to the exchange interaction, the geometrically broken symmetry results in the reorganization of the magnetostatic energy terms in the form, adapted to the geometry: $E_{\text{d}} = E_\text{ms}^{0} + E_{\text{d}}^{\textsc{a}} +  E_{\text{d}}^{\textsc{c}} + E_{\text{d}}^{{\textsc{s-v}}}$. The term $E_\text{ms}^{0}$ is similar to the planar case, 
\begin{subequations} \label{eq:E-ms-nonloc}
\begin{equation} \label{eq:E0-ms}
\begin{aligned}
E_\text{ms}^{0} &= \frac{1}{8\pi} \int \vec{m}(\vec{r}) \cdot \mathrm{d} \vec{S}  \int  \frac{\vec{m}({\vec{r}'})\cdot \mathrm{d} {\vec{S}'}}{\left|\vec{r} - {\vec{r}'}\right|}\\
& + \frac{1}{8\pi} \int \mathrm{d} \vec{r}\ \eth_\alpha m_\alpha(\vec{r}) \int \mathrm{d} {\vec{r}'} \frac{\eth_\alpha m_\alpha({\vec{r}'})}{\left|\vec{r} - {\vec{r}'}\right|}.
\end{aligned}
\end{equation}  

Here, $\mathrm{d}\vec{S} = \vec{n}^\text{+~(-)}\mathrm{d}S$ is a directed surface element.
In the main order on the shell thickness $h$, the above magnetostatic energy term  is $E_\text{ms}^0 = (h/2)\int (\vec{m}\cdot \vec{\hat{n}})^2 \mathrm{d}S + \mathcal{O}(h^2)$ \cite{Carbou01,Slastikov05,Fratta16b}. This term is local and typically leads to the renormalization of anisotropy coefficients. It is the only term, linear in $h$ stemming from the magnetostatic interaction.
All other contributions to the magnetostatic interaction are essentially nonlocal. In thin shell limit they scale as $h^2 + \mathcal{O}(h^3)$.

The next magnetostatic term reads
\begin{equation} \label{eq:E1-ms}
\begin{aligned}
E_{\text{d}}^{\textsc{a}} = & \frac{1}{4\pi}\! \int w_\text{d}^{\textsc{a}} \mathrm{d} \vec{r},\\
 w_\text{d}^{\textsc{a}} = & \mathfrak{S}(\vec{r}) \left[ \frac12 \int \frac{\mathfrak{S}({\vec{r}'})\mathrm{d} {\vec{r}'}}{\left|\vec{r} - {\vec{r}'}\right|} - \int \frac{\vec{m}({\vec{r}'}) \cdot \mathrm{d} {\vec{S}'}}{\left|\vec{r} - {\vec{r}'}\right|} \right].
\end{aligned}
\end{equation} 
Although nonlocal, this term is bilinear on the normal component of magnetization and contributes to the shape anisotropy.

The curvature-induced chiral part of the nonlocal magnetostatic interaction reads
\begin{equation} \label{eq:E2-ms}
E_{\text{d}}^{\textsc{c}} \! = \! \frac{1}{4\pi} \!\!\int \!\! w_\text{d}^{\textsc{c}} \mathrm{d} \vec{r},\quad  w_\text{d}^{\textsc{c}} = -\eth_\alpha m_\alpha(\vec{r}) \!\int\! \frac{\mathfrak{S} ({\vec{r}'}) \mathrm{d} {\vec{r}'}}{\left|\vec{r} - {\vec{r}'}\right|}\!.\!\!
\end{equation} 
It characterizes the interaction between `common' volume charge $ \eth_\alpha m_\alpha(\vec{r}) $ and the curvature-induced charge $\mathfrak{S}$. Thus, the energy \eqref{eq:E2-ms} is specific to curved shells only. Similarly to the curvature-induced DMI $E_{\text{x}}^\textsc{d}$, the magnetostatic contribution $E_{\text{d}}^{\textsc{c}}$ is linear with respect to the derivative of magnetization. Having a similarity with the Lifshitz invariants in Eq.~\eqref{eq:Eex-D}, this energy term favours the coupling between the out-of-surface magnetization $m_n$ and spatial derivatives of the in-surface components $m_\alpha$. Therefore, this term is responsible for nonreciprocal effects, in particular, magnetochiral effects. We emphasize that in contrast to the curvature-induced DMI \eqref{eq:Eex-D}, this chiral term~\eqref{eq:E2-ms} is essentially nonlocal.

The last term in magnetostatics describes the interaction between surface and volume magnetostatic charges:
\begin{equation} \label{eq:E3-ms}
E_{\text{d}}^{{\textsc{s-v}}} \!=\! \frac{1}{4\pi} \!\! \int \!\! w_\text{d}^{\textsc{s-v}} \mathrm{d} \vec{r},\quad w_\text{d}^{\textsc{s-v}} \!=\! \eth_\alpha m_\alpha(\vec{r}) \!\int\! \frac{\vec{m}({\vec{r}'}) \!\cdot\! \mathrm{d} {\vec{S}'}}{\left|\vec{r}- {\vec{r}'}\right|}\!.\!\!
\end{equation}
This coupling between surface and volume magnetostatic charges does not exist in planar fimls.
It also vanishes for any homogeneous magnetic texture in curved shell. We point out the interaction \eqref{eq:E3-ms} is chiral and appears if the top and bottom surfaces of a shell are not equivalent, i.e. they cannot be translated one into another by translation along the normal. As a stark consequence, novel chiral effects emerge in spatially corrugated magnetic thin films. For instance, this term appears in cylindrical and spherical shells due to the difference in the area of the inner and outer surfaces.
\end{subequations}

The energy of intrinsic DMI is broken into two componets: $E_\textsc{dm} = E_\textsc{dm}^0 + E_\textsc{dm}^\textsc{a}$~\cite{Kravchuk16a}. Here, $E_\textsc{dm}^0$ is a regular part of DMI with a structure, similar to the planar case:
\begin{subequations}\label{eq:intrinsic-DMI}
\begin{equation}\label{eq:DMI-0}
E_\textsc{dm}^0 = hD \int  \left( m_n \eth_{\alpha}m_\alpha - m_\alpha \eth_{\alpha} m_n \right) \mathrm{d}S,
\end{equation}
cf.~Eq.~\eqref{eq:Eex-D}. The second part plays a role of an additional uniaxial anisotropy
\begin{equation}\label{eq:DMI-a}
E_\textsc{dm}^\textsc{a} = - hD \int \mathfrak{S}m_n \mathrm dS,
\end{equation}
cf.~Eq.~\eqref{eq:E1-ms}. 
\end{subequations}

%%%%%%%%%%%%%%%%%%%%%%%%%%%%%%%%%%%%%%%%%%%%%%%%%%
%%%			New section
%%%%%%%%%%%%%%%%%%%%%%%%%%%%%%%%%%%%%%%%%%%%%%%%%%

\section{Discussion}
\label{sec:discussion}

\begin{table*}
\caption{\label{tbl:classif-ground} Effects of curvature-induced chiral and anisotropic terms in exchange and magnetostatic energies on \emph{assumed equilibrium state} $\widetilde{\vec{m}}$ given by the anisotropy for different symmetries. Here, each of curvatures is considered either zero or nonconstant function. The central and right hand parts of the table contains consequences of the input given in the left. Abbreviations EA HA correspond to easy-axis and hard axis anisotropies respectively. Last five columns show presence of geometry-induced exchange and magnetostatic terms. Black arrows and dotted lines in surfaces show principal directions for the corresponding surfaces. A direction $\vec{e}_a$ is one of the unit vectors or their linear combination.}
\includegraphics[width=180mm,trim={2mm 72mm 2mm 71mm},clip]{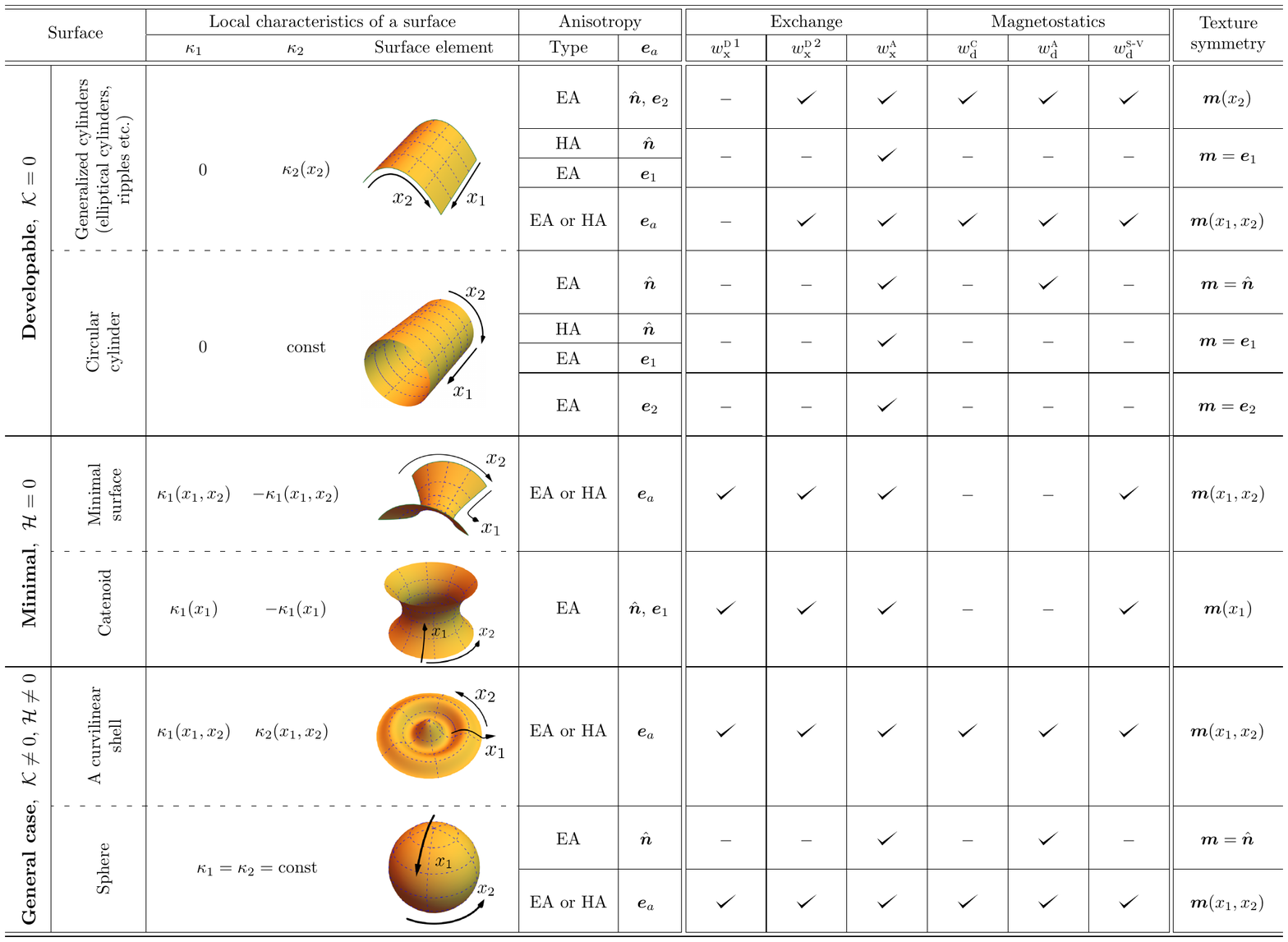}
\end{table*}

The direct analysis of all magnetostatic energy contributions~\eqref{eq:E-ms-nonloc} is  complicated by the nonlocal integration kernels. For this reason, we apply a symmetry analysis to the energy of a ferromagnetic shell to distinguish sources of possible effects of curvature on the magnetic texture. 

In the following, we consider the case of strong anisotropies, which allows us to study a magnetic texture, which does not deviate significantly from the \emph{assumed equilibrium state} $\widetilde{\vec{m}}$ given by the anisotropy. We are interested in how local properties, i.e. local curvatures of the surface and local orientation of the magnetic easy axis, impact the resulting global magnetic state. 

%%%%%%%%%%%%%%%%%%%%%%%%%%%%%%%%%%%%%%%%%%%%%%%%%%%
%%%%			New subsection
%%%%%%%%%%%%%%%%%%%%%%%%%%%%%%%%%%%%%%%%%%%%%%%%%%%
%
%
\subsection{Effects of curvature, classified by the shell type}
\label{sec:classification}

Any surface $\vec{\varsigma}$ can be locally defined via its two principal curvatures $\kappa_1$ and $\kappa_2$, which are present in the energy terms discussed above. For our discussion, we consider uniaxial magnets with special types of anisotropy along one of the principal directions for the following distinct cases of surfaces: 

(i) A class of developable surfaces of zero Gaussian curvature, $\mathcal{K}(\vec{r}) = 0$, includes cylinders, cones and tangent surfaces~\cite{Krivoshapko15}. They can be locally developed into a plane without stretching. Since cones and tangent surfaces are singular ones~\cite{Johansen05}, here, we consider generalized cylindrical surfaces only.

(ii) Minimal surfaces with vanishing mean curvature, $\mathcal{H}(\vec{r}) = 0$, have principal curvatures of opposite signs and in the vicinity of each point they are saddle-shaped. Minimal surfaces provide the minimal surface area enclosed by a given boundary. 

(iii) General case with nonvanishing $\mathcal{H}$ and $\mathcal{K}$ and arbitrary local surface elements including convex and saddle ones. 

%(iv) Furthermore, it is insightful to distinguish a subclass of surfaces of revolution. In the following, we consider easy-normal, easy-surface and easy-axis anisotropies along each of the principal directions.

The impact of the geometry on a magnetic texture is summarized in Table~\ref{tbl:classif-ground}. It is given by the interplay of the curvature-induced energy terms and the type of anisotropy and orientation of the anisotropy axis. We refer to the curvature-related energy terms as following:
\begin{equation}\label{eq:}
\begin{aligned}
E = & E_0 + 2 h\ell^2 \int \mathrm dS \left( w_\text{x}^{\textsc{d}\,1} + w_\text{x}^{\textsc{d}\,2} + w_\text{x}^{\textsc{a}} \right) \\
    & + \dfrac{1}{4\pi} \int \mathrm d\vec{r} \left( w_\text{d}^{\textsc{c}} + w_\text{d}^{\textsc{a}} \right).
\end{aligned}
\end{equation}
Here $E_0 = E_\text{ex}^0 + E_\text{ms}^{0} + E_{\text{d}}^{{\textsc{s-v}}}$ absorbs terms which do not depend explicitly on the curvature of the surface. First two of them contain derivatives of magnetization components and can result in chiral effects even in a purely planar case due to chiral magnetic texture, the so-called \emph{pattern-induced chirality breaking} ~\cite{Streubel16a}. These effects are well studied for magnons on the background of solitons \cite{Sheka01}, vortices \cite{Sheka04} and skyrmions \cite{Kravchuk18}. We do not discuss influence of the intrinsic DMI~\eqref{eq:intrinsic-DMI} here as it has symmetry of already included terms.

Third term, $E_{\text{d}}^{{\textsc{s-v}}}$, is present only for the inhomogeneous magnetization texture if top and bottom surfaces of the shell are not equivalent.

The curvature-induced exchange terms $w_\text{x}^{\textsc{d}\,1}$, $ w_\text{x}^{\textsc{d}\,2}$ and $w_\text{x}^{\textsc{a}}$ scale linearly with the shell thickness. Both magnetostatic terms, $ w_\text{d}^{\textsc{c}} $ and $w_\text{d}^{\textsc{a}} $
are present only for curved shells with a non-zero mean curvature, $\mathcal{H}\neq 0$; for the infinitesimally thin shells they scale quadratically with thickness. These magnetostatic terms are absent for minimal surfaces, e.g. for catenoids and helicoids.

%==================================================================\
\begin{figure*}
\includegraphics[width=\linewidth]{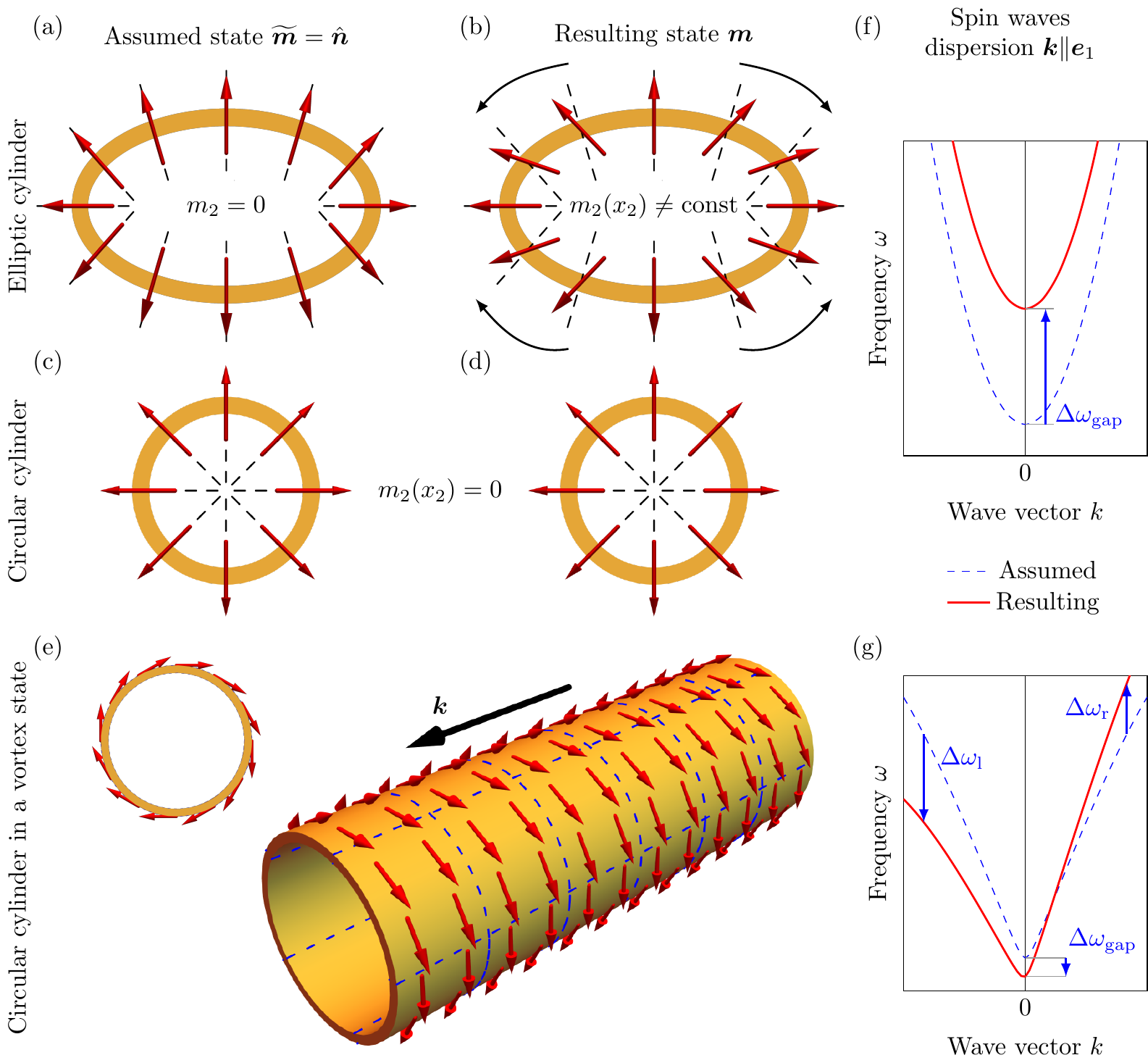}
\caption{(Color online) \textbf{Schematics of modification of the \emph{assumed equilibrium state} and spin wave spectrum by curvature.} (a)--(d) Elliptical and circular cylinders with easy-normal anisotropy affected by curvature in different ways: Magnetization pattern in the elliptical cylinder is modified due to the symmetry breaking. Magnetization is shown by red arrows, normal direction is shown by dashed lines. (e) Static vortex state in the circular cylinder is modified by $w_\text{x}^a$ only. (f),~(g) Dispersion of spin waves propagating along cylinder axis $\vec{e}_1$ [is shown by black arrow in panel (e)]. While dispersion curve is only shifted by $\Delta \omega_{\text{gap}}$ for easy-normal anisotropy due to contribution of $w_\text{x}^\textsc{a}$, also it becomes asymmetric for the vortex state due to $w_\text{d}^\textsc{c}$ and $w_\text{d}^\textsc{s-v}$~\cite{Otalora16,Otalora17}. Blue arrows show change from assumed to the actual dispersion curve with $\Delta \omega_{\text{l}}$ and $\Delta \omega_{\text{r}}$ being frequency shifts for the left and right branches of the lowest radially symmetric mode respectively. }
\label{fig:influence}
\end{figure*}
%==================================================================/

It is important to stress that such an approach can not be considered as a sufficient condition of existence and moreover stability of corresponding magnetization states. 
 
Table~\ref{tbl:classif-ground} provides the following information. If we know local curvatures and the direction of the easy axis in the vicinity of a given point, then we can assess if the resulting magnetic texture will be modified due to the presence of a curvature-induced anisotropy and if the texture will be chiral. We consider that the respective energy term will impact the texture if the term is nonzero. \emph{No other criteria are applied} while assembly Table~\ref{tbl:classif-ground}. In particular, this Table cannot be considered as sufficient conditions of existence and stability of the magnetization state. Here we discuss possible statical states by applying only symmetrical arguments to the energy functional  irrelevant they are in local minimum of energy or not. Investigation of equilibrium magnetization texture for the concrete geometry should be a purpose of a separate work. For example, $w_\text{x}^{\textsc{d}\,1}$ does not impact magnetic textures (it vanishes) for the following cases: either $\kappa_1 \equiv 0$, or magnetic texture does not vary  along $x_1$, or $m_n \equiv 0$. 
A possible magnetic texture is assumed based on the sample symmetry and interplay between intrinsic and curvature-induced anisotropies. 

For instance, we consider a developable surface with $\mathcal{K} = 0$ and non-constant second principal curvature, e.g., elliptical cylinder or ripple, and assume that magnetic easy axis is pointing normally to the surface. Then, the magnetic texture $\vec{m}= \vec{m}(x_2)$ is influenced by $w_\text{x}^{\textsc{d}\,2}$ (local chiral term). Both chiral and anisotropic magnetostatic terms, $w_\text{d}^\textsc{c}$ and $w_\text{d}^\textsc{a}$, respectively, are also present since the mean curvature is nonzero (these terms are essentially nonlocal). The term $w_\text{d}^\textsc{s-v}$, which is responsible for the interaction between the surface and volume charges, can appear due to inequivalence of top and bottom surfaces of the shell for inhomogeneous magnetization texture with non-vanishing `common' volume magnetostatic charges. Based on these considerations, the assumed equilibrium state $\widetilde{\vec{m}}$ (e.g. normally magnetized elliptical cylinder) will be modified due to local and nonlocal curvature effects as follows: (i) the state will be chiral, i.e. deviation from the $\hat{\vec{n}}$ is linear with respect to $\kappa_2(x_2)$, and (ii) effective easy-normal anisotropy will be inhomogeneously changed. As a result of this consideration, the initially assumed strictly normal magnetization distribution is modified by the appearance of the $m_2$ component: $\vec{m}= \{0, m_2(x_2), m_n(x_2)\}$. 

Minimal surfaces do not exhibit effects from magnetostatics, which explicitly depend on the curvature due to $\mathcal{H} \equiv 0$. At the same time, all geometry-induced exchange-driven terms are present for any texture symmetry except easy-surface anisotropy or easy-axis anisotropy along $\vec{e}_2$. The chiral magnetostatics-driven term $w_\text{d}^{\textsc{s-v}}$ is always present for inhomogeneous textures with non-zero magnetostatic charge if the top and bottom surfaces of a shell are not equivalent. As in the previous case, the initially assumed strictly normal magnetic texture is modified by the appearance of the $m_1$ component: $\vec{m}= \{m_1(x_1), 0, m_n(x_1)\}$ for a catenoid.

For the general case of $\mathcal{H} \neq 0$ and $\mathcal{K} \neq 0$, any texture is expected to become chiral and modified due to the curvature-induced anisotropy (local and nonlocal). Note, that Table~\ref{tbl:classif-ground} is also valid for nonlinear excitations of the equilibrium state like domain walls if their symmetry corresponds to the function given in the `Texture symmetry' column.

%%%%%%%%%%%%%%%%%%%%%%%%%%%%%%%%%%%%%%%%%%%%%%%%%%%
%%%%			New subsection
%%%%%%%%%%%%%%%%%%%%%%%%%%%%%%%%%%%%%%%%%%%%%%%%%%%
%
%
\subsection{Special cases of magnetic shells}

The special interest attracts the spherical geometry with $\kappa_1 = \kappa_2 = \text{const}$, see Table~\ref{tbl:classif-ground}. The curvature-induced exchange driven DMI \eqref{eq:Eex-D} is well-established for the spherical surfaces. For magnetic vortices it results in coupling between the localized out-of-surface vortex core structure $m_n$ and the delocalized in-surface magnetization texture $m_1$, the so-called polarity chiralty coupling \cite{Kravchuk12a}. Note that without nonlocal magnetostatic interaction the magnetization states on the sphere forms three dimensional onion state with the in-surface meridional direction. Very recently it was shown that the volume magnetostatics  results in a whirligig state \cite{Sloika16c}, which has no `common' volume charges, hence the magnetostatic energy of such a state is described by curvature-induced charges $\mathfrak{S}=2m_n/R$, see \eqref{eq:E1-ms}.

Using a spherical shell with easy-surface anisotropy as a reference example, let us estimate conditions of nonlocality of curvature effects. We consider a shell with localized curvature in a shape of some curved bump accommodating a localized topological defect. The defect size $w$ is much smaller than the typical curvature radius. Under this assumption, the curvature can be assumed constant (as for sphere, $\kappa_1(0) = \kappa_2 (0) =\kappa_0$) in the vicinity of the magnetic defect, hence we model the surface near the defect as the spherical one. The local curvature effects are determined mainly by the curvature-induced DMI, $E_{\text{x}}^\textsc{d}\sim h \ell^2\kappa_0 w$. The nonlocal curvature effects are mainly caused by the volume magnetostatic charges. Using the asymptotic analysis similar to \cite{Sloika17}, one can estimate that in the main order on the curvature, the magnetostatic contribution is determined by $ h^2 w^2 \kappa_0$. Both energy terms, which describe curvature effects, local and nonlocal ones, become of the same order when the film thickness is $h_c\approx \ell^2/w$. For topological defects with a typical size $w$ similar to the exchange length $\ell$, we obtain $h_c\sim \ell$. Note that the defect size can be much smaller than the exchange length: e.g., the curvature-induced skyrmion in a spherical shell with easy-normal anizotropy has a typical size $w\sim \ell^2\kappa_0$ \cite{Kravchuk16a}, which results in $h_c\sim 1/\kappa_0\gg \ell$. For thin films with $h\lesssim h_c$ the local picture with the exchange-driven curvature effects is adequate. Thicker films require nonlocal effects to be considered in the description of magnetic textures.

Developable surfaces are a special case due to the absence of $w_\text{x}^{\textsc{d}\,1}$ term including a family of circular and elliptical cylinders. The effect of curvature on magnetic state can be illustrated for the case of strong easy-normal anisotropy. According to Table~\ref{tbl:classif-ground}, the assumed normal state $\widetilde{\vec{m}}=\hat{\vec{n}}$ is affected by the following chiral contributions, $ w_\text{x}^{\textsc{d}\,2}$, $w_\text{d}^{\textsc{c}}$, and $ w_\text{d}^{\textsc{s-v}}$ in the case of elliptic cylinder, see Figs.~\ref{fig:influence}(a)--(b). The tilt of magnetization from the normal direction is mainly determined by the curvature variation, $\partial_2\kappa_2$. Analogous considerations allows to conclude that the assumed state along the normal direction will remain for the circular cylinder with $\kappa_2=\text{const}$, see Figs.~\ref{fig:influence}(c)--(d).

We focused the above analysis on simple magnetization texture. Nevertheless, the proposed theory can be used to describe also complex static and dynamic excitations.
To illustrate this approach we consider the spin wave propagation along straight generatrix $\vec{e}_1$ on the equilibrium state of cylinders, obtained in Fig.~\ref{fig:influence}. The analysis of the energy terms predicts the reciprocal propagation of spin waves in cylinders with easy-normal anisotropy for both elliptic and circular cases. Similar effects are known for the planar films with intrinsic DMI \cite{Cortes-Ortuno13}. The curvature-induced anisotropy $w_{\text{ex}}^\textsc{a}$ as well as magnetostatic term $w_\text{d}^{\textsc{a}}$ are responsible for the shift of the magnon gap $\Delta \omega_{\text{gap}}$, see Fig.~\ref{fig:influence} (f). Note this effect is similar to the magnon gap shift for the vortex domain wall in circular cylinders \cite{Gonzalez10}.

Now we consider the circular cylinder with anisotropy along $\vec{e}_2$. According to Table~\ref{tbl:classif-ground}, there are no chiral effects for the equilibrium state, and the resulting vortex state corresponds to the azimuthal anisotropy direction. One can excite the spin waves propagating along $\vec{e}_1$. They are influenced by curvature-induced anisotropy $w_\text{x}^\textsc{a}$, which results in magnon gap shift $\Delta \omega_{\text{gap}}$, see Fig.~\ref{fig:influence}(g). Typically, the analysis of energy terms is insufficient for the description of dynamical excitations such as spin waves: one needs to analyse Landau-Lifshitz equations. The key point of such analysis is to derive an effective field. In the case under consideration the only nonvanishing chiral energy terms can be written through the normal components of the effective dipolar fields $H_n^\textsc{c}$ and $H_n^\textsc{s-v}$:
\begin{equation} \label{eq:E-ms-cylinder}
\begin{aligned} 
\!w_\text{d}^{\textsc{c}\,\text{cyl}} & = - m_n(x_1) H_n^\textsc{c}\\
 &  =- \dfrac{m_n(x_1)}{\rho} \int  \frac{ \partial_1' m_1(x_1') \mathrm{d} \vec{r'}}{\left|\vec{r} - \vec{r'}\right|},\\
\!w_\text{d}^{\textsc{s-v}\,\text{cyl}}\! & = - m_n(x_1)H_n^\textsc{s-v} \\
& =\! \rho m_n(x_1)  \!\!\int \!\!\mathrm{d} x_1' \mathrm{d} x_2'  \frac{\partial_1' m_1(x_1')}{\left|\vec{r} - \vec{r'}\right|}\Biggr\rvert_{\rho=R}^{\rho=R+h} \!\!,
\end{aligned}
\end{equation}
with $\rho$ being the cylindrical radius. One can see, that the dynamical magnetization $m_1\propto e^{i k_1 x_1}$ results in the break of the mirror invariance $x_1 \to -x_1$ of the energy densities as well as effective fields.
Finally, the wave length of magnons at a given frequency is different for opposite propagation directions, which results in a splitting of the spin wave states with  left- and right handed chiralities, which was very recently studied in Refs.~\cite{Otalora16,Otalora17}, see Fig.~\ref{fig:influence}(g).

%%%%%%%%%%%%%%%%%%%%%%%%%%%%%%%%%%%%%%%%%%%%%%%%%%
%%%			New section
%%%%%%%%%%%%%%%%%%%%%%%%%%%%%%%%%%%%%%%%%%%%%%%%%%

\section{Conclusions and outlook}
\label{sec:conclusions}

The magnetism in curved geometries encompasses a range of fascinating geometry-induced effects in the magnetic properties of materials \cite{Streubel16a}. Here we propose a platform for theoretical analysis of magnetization textures in curvilinear ferromagnetic shells of different geometries. The developed generalized micromagnetic theory of curvilinear ferromagnetic shells allows to treat together both local (exchange and anisotropy) and non-local (magnetostatics) interactions. 

To illustrate our theory we classify possible curvature effects on the equilibrium states.
We focus our analysis on rather simple magnetization structures, mostly defined by the anisotropy. Nevertheless, the developed theory is general: it allows to describe also strongly nonlinear magnetization texture, e.g., domain walls, vortices, skyrmions. It is important to specify that our illustrations of proposed micromagnetic theory of curvilinear ferromagnetic shells are based only on symmetrical arguments for the energy functional. In particular, these arguments cannot be considered as sufficient conditions of existence of magnetization state. Nevertheless, we hope that the presented work will open a pull for further investigations of magnetization textures for the concrete geometries. In particular, the proposed theory can be applied for the prediction of properties and responses of curved thin films. This allows to carry out targeted design and optimization for specific spintronic and magnetooptic devices and applications. The proposed theory can be generalized to include intrinsic Dzyaloshinskii-Moriya interaction of the film using by introducing mesoscale DMI \cite{Volkov18}. Still, the key impact of the developed theory is in the possibility to tailor the properties of `standard' ferromagnets to realize chiral textures.
These developments will pave the way towards new device ideas relying on curvature effects in magnetic nanostructures.

We do expect that the impact goes well beyond the magnetism community. The presented conclusions can be easily extended for studying the evolution of generic vector fields on curved shells in different models of condensed (graphene \cite{Yan13a}, superconductors \cite{Vitelli04}) and soft (nematics, cholesterics \cite{Napoli12}) matter.

\begin{acknowledgements}
The authors thank Dr. Volodymyr Kravchuk for helpful discussions. D.~D.~S. and O.~V.~P.  thank Helmholtz-Zentrum Dresden-Rossendorf e.~V. (HZDR), where part of this work was performed, for their kind hospitality and acknowledge the support from the Alexander von Humboldt Foundation (Research Group Linkage Programme). O.~V.~P. acknowledges the support from DAAD (code No. 91530902). We acknowledge the support from FONDECYT 1161403 and Centers of excellence with Basal/CONICYT financing, grant FB0807, CEDENNA. This work was supported by the Program of Fundamental Research of the Department of Physics and Astronomy of the National Academy of Sciences of Ukraine (Project No. 0116U003192). This work was partially supported by Taras Shevchenko National University of Kyiv (Projects 19BF052-01 and 18BF052-01M).
\end{acknowledgements}

\appendix
%
%
%%%%%%%%%%%%%%%%%%%%%%%%%%%%%%%%%%%%%%%%%%%%%%%%%%%
%%%%			New section
%%%%%%%%%%%%%%%%%%%%%%%%%%%%%%%%%%%%%%%%%%%%%%%%%%%
%

\section{Darboux three-frame and modified covariant derivatives}
\label{sec:geometry}

In order to describe magnetic properties of the curved shell we choose the curvilinear reference frame adapted to the geometry. Then the spatial variation of the anisotropy axes is automatically accounted for, and the anisotropy energy density assumes its usual translation-invariant form. To be more specific, we define principle curvatures $\kappa_1$ and $\kappa_2$, which are maximum and minimum of the normal curvature at a given point on a surface \cite{Weisstein03}. Then, we can define the directions $\vec{e}_1$ and $\vec{e}_2$, in which the principle curvatures occur, the so-called, principle directions \cite{Weisstein03}. Now we construct an orthonormal Darboux three-frame $\left\{\vec{e}_1, \vec{e}_2, \vec{n}\right\}$ on the surface $\vec{\varsigma}$, where $\vec{n} = \vec{e}_1\times \vec{e}_2$ is the normal to the surface \cite{Chern99}. The local curvilinear coordinates $\left\{x_1, x_2\right\}$ correspond to lines of curvatures; the unit vectors $\vec{e}_\alpha = \partial_\alpha \vec{\varsigma}/|\partial_\alpha \vec{\varsigma}|$. Here and in what follows, we use Greek letters $\alpha,\,\beta,\ldots = {1,2}$ to denote indices restricted to the shell surface; to indicate all three components of some vector we use the Latin indices  $i,\,j,\ldots = {1,2,3}$. The first fundamental form (surface metric) $g_{\alpha\beta} = \partial_\alpha \vec{\varsigma} \cdot \partial_\beta \vec{\varsigma}$. The local curvilinearity is determined by the second fundamental form tensor $b_{\alpha \beta} = \vec{\hat{n}} \cdot \partial^2_{\alpha, \beta} \vec{\varsigma}$. In the local surface reference frame the Weingarten map $|| H_{\alpha \beta} || =  ||b_{\alpha \beta}/\sqrt{g_{\alpha \alpha}g_{\beta \beta}}||$ has  a simple diagonal form, $||H_{\alpha \beta}|| = \text{diag}(\kappa_1,\kappa_2)$ with $\kappa_\alpha$ being the principal curvature. The Gaussian curvature $\mathcal{K}=\kappa_1 \kappa_2$ and mean curvature $\mathcal{H}=\kappa_1+\kappa_2$.

\begin{figure}[hb]
	\includegraphics[width=\linewidth]{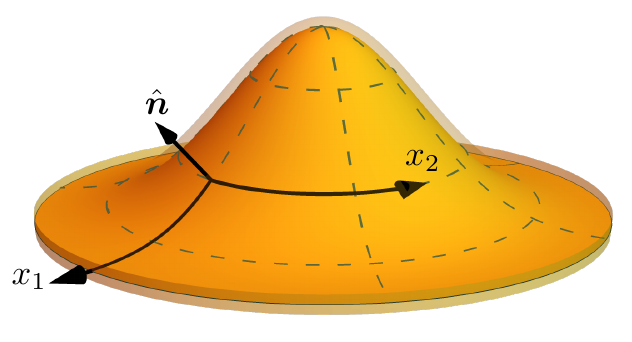}
	\caption{(Color online) \textbf{Schematics of reference frame construction on the surface of revolution.} Curved arrows with $x_1$ and $x_2$ labels show principal directions. The $x_2$ direction is undefined in the apex.}
	\label{fig:surf-rev}
\end{figure}

Let us parametrize the ferromagnetic shell using the thin-shell limit; we define a finite thickness shell (by extruding surface $\vec{\varsigma}$ in the normal direction),
$\vec{r}(x_1,x_2,x_3) = \vec{\varsigma}\left(x_1,x_2\right) + x_3 \hat{\vec{n}}$, where $x_3\in[-h/2,h/2]$ is a cross--section (thickness) coordinate. According to Dupin's theorem~\cite{Weisstein03} the the reference frame $\left\{x_1, x_2, x_3\right\}$ is orthogonal with $\vec{\varsigma}$ being the coordinate isosurface $x_3=\text{const}$.
Schematic of the reference frame for the particular case of the revolution surface is plotted in Fig.~\ref{fig:surf-rev}.

The exchange energy of a classical ferromagnet, $E_{\text{x}} =- \ell^2 \int \mathrm{d} \vec{r}\, (\vec{m}\cdot \vec{\nabla}^2\vec{m})$, can be treated in thin shell limit, when the magnetization does not changes in the transversal direction, $\vec{m} = \vec{m}\left(x_1,x_2\right)$. By applying a surface Laplacian  in its curvilinear form reads $\vec{\nabla}^2 = (1/\!\sqrt{g}) \partial_\alpha \left( \sqrt{g} g^{\alpha\beta} \partial_\beta \right)$ with $ g = \det \|g_{\alpha\beta}\|$ and dual basis $\|g^{\alpha\beta}\| = \|g_{\alpha\beta}\|^{-1}$, one can restructure the exchange energy to the form \eqref{eq:E-ex}, adapted to the curvilinear geometry.

The bullet point of such reorganization is a covariant derivative apparatus. The (modified) covariant derivative is defined as follows:
\begin{equation} \label{eq:D-covar}
\begin{aligned}
\eth_\alpha m_\beta &:= \frac{\partial_\alpha m_\beta + \epsilon_{\beta\gamma} \left(\vec{e}_1\cdot \partial_\alpha  \vec{e}_2 \right)m_\gamma}{\sqrt{g_{\alpha\alpha}}},\\ 
\eth_\alpha m_n &:= \frac{\partial_{\alpha} m_n}{\sqrt{g_{\alpha\alpha}}}.
\end{aligned}
\end{equation}
Such a definition coincides (up to the factor $\sqrt{g_{\alpha\alpha}}$) with the standard definition of the covariant derivative for the tangential vector components. For convenience, we introduce similar notation for the normal vector component.

The main purpose to use the language of covariant derivatives is to separate two effects: (i) an explicit curvature effect and (ii) spurious effect of the curvilinear reference frame. To illustrate the difference let us consider the flat film, where curvature effects are absent. Using a polar reference frame $(x_1, x_2) = (\rho,\chi)$ one can obtain the following exchange energy density:
\begin{equation} \label{eq:exch-polar}
\begin{aligned}
w_\text{x} & = \frac{1}{g_{\alpha\alpha}} (\partial_\alpha m_i) (\partial_\alpha m_i) \\
& + \dfrac{m_1^2 + m_2^2}{\rho^2}  + \dfrac{2}{\rho^2} \left(m_1 \partial_2 m_2 - m_2 \partial_2 m_1 \right)
\end{aligned}
\end{equation}
with metric tensor $\|g_{\alpha\beta}\| = \operatorname{diag}(1, 1/\rho)$. While the first term in~\eqref{eq:exch-polar} can have the structure typical for the isotropic exchange interaction in planar system, last two terms can be misinterpreted as some anisotropy and DMI. Moreover, these two spurious terms formally diverge at origin as a direct consequence of the polar reference frame apparatus. Unlike this coordinate dependent presentation, the covariant formulation of the exchange interaction \eqref{eq:E-ex}
\begin{equation} \label{eq:covariant-polar}
w_\text{x} = \left(\eth_\alpha m_i\right) \left(\eth_\alpha m_i\right),
\end{equation}
is free of spurious terms, which mimic the curvature-induced effects.

%==================================================================\
%==================================================================/

%
%----------------------------------------------------------------
%
%\bibliography{soliton}

\end{document}